\documentclass[prb,amsmath,twocolumn,showpacs]{revtex4-1}
\usepackage{bm}
\usepackage{amssymb}
\usepackage{colordvi}
\usepackage{graphicx}
\usepackage{color}
\usepackage{hyperref}

\newcommand{\be}{\begin{equation}}
\newcommand{\ee}{\end{equation}}
\newcommand{\bea}{\begin{eqnarray}}
\newcommand{\eea}{\end{eqnarray}}

\def\nn{\nonumber}

\begin{document}

\title{Enhancing efficiency and power of quantum-dots resonant tunneling 
  thermoelectrics in three-terminal geometry by cooperative effects}

\date{\today}

\author{Jian-Hua Jiang}\email{jianhua.jiang.phys@gmail.com}
\affiliation{Department of Physics, University of Toronto, Toronto,
  Ontario, M5S 1A7 Canada}
\affiliation{Center for Phononics and Thermal Energy Science, School
  of Physical Science and Engineering, Tongji University, Shanghai, 200092 China}

\begin{abstract}
We propose a scheme of multilayer thermoelectric engine where
{\em one} electric current is coupled to {\em two} temperature gradients in
three-terminal geometry. This is realized by resonant tunneling through
quantum dots embedded in two thermal and electrical resisting polymer matrix
layers between highly conducting semiconductor layers. There are two thermoelectric
effects, one of which is pertaining to inelastic transport processes
(if energies of quantum dots in the two layers are different)
while the other exists also for elastic transport processes. {These two
correspond to the transverse and longitudinal thermoelectric effects
respectively and are associated with different temperature gradients.} We show
that cooperation between the two thermoelectric effects leads to 
markedly improved figure of merit and power factor {which is confirmed
by numerical calculation using  material parameters.
Such enhancement is robust against phonon heat conduction and energy
level broadening. Therefore we demonstrated cooperative
effect as an additional way to effectively improve performance of
thermoelectrics in three-terminal geometry.}
\end{abstract}

% key words: thermoelectric energy conversion, nanostructured
% thermoelectrics, three-terminal thermoelectric devices, collective
% thermoelectric effect, inelastic transport processes


\maketitle

\section{Introduction}

Harvesting usable energy from wasted heat using thermoelectrics has
been attracting a lot of research interest.\cite{honig} Much efforts
have been devoted to bulk materials, making them mature thermoelectric
systems that are already useful in industrial technologies.\cite{rev,rev2} 
Recently there is a trend of incorporating nanostructures to further
improve the performance of thermoelectric materials.\cite{nano1,nano2,lowd}
Successful examples have been achieved in many materials/structures along this
direction.\cite{nano1,nano2,lowd} Theoretical studies have demonstrated
that the thermoelectric properties of individual nanostructures can be
much better than the bulk.\cite{joe,hicks,ms,Linke} Experimental
efforts have pushed forward the measurements of thermoelectric
properties of individual nanostructures.\cite{nano-exp,nano-th} There are also
studies trying to fill the gaps between the thermoelectric properties
of individual nanostructures and their assemblies\cite{nano1,nano2,lowd} as
well as attempts to improve thermoelectric performance by tuning the
shape and organization patterns of nanostructures.\cite{nano-ordering}

\begin{figure}[htb]
  \includegraphics[height=7.9cm]{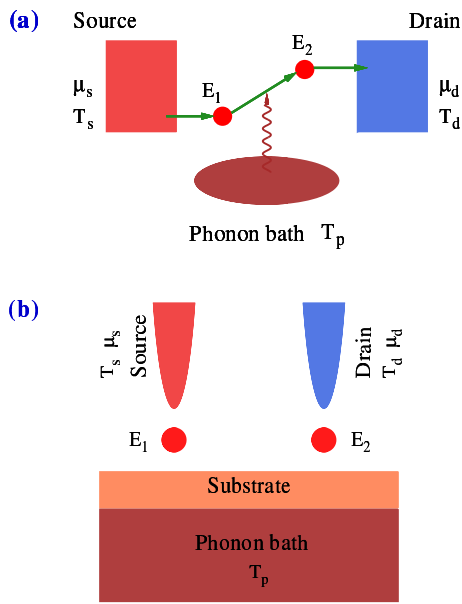}
  \caption{ (Color online) Illustration of a three-terminal
    thermoelectric system (studied in Ref.~\onlinecite{prb1}). (a) An
    electron first tunnels from the 
    source to the quantum dot (represented by a red dot)
    with energy $E_1$ and then hops to the other quantum 
    dot with energy $E_2$ by absorbing a phonon from the phonon
    bath. After that the electron tunnels into the drain
    electrode. The tunneling processes are elastic, while the hoping
    between quantum dots must be assisted by a phonon with energy
    $E_2-E_1>0$. The temperatures of the source, drain and phonon bath
    are $T_s$, $T_d$, and $T_p$, while the electrochemical 
    potentials of the source and drain are $\mu_s$ and $\mu_d$, 
    respectively. When the phonon bath has the highest temperature
    among the three reservoirs,
    excess number of phonons prefer phonon absorption processes and
    hence induces an electric current from the source to the
    drain. (b) In realistic situations the phonon bath is connected
    with the (insulating) substrate supporting the quantum dots while the two
    electrodes are suspended to be isolated from the phonon bath.}
\label{fig1}
\end{figure}

In addition to  material and structural aspects, geometry also
plays an important role in thermoelectric applications. For example, 
transverse thermoelectrics\cite{TTE} take advantages of
accumulating temperature difference in one direction while
generating electric current in the perpendicular direction. Geometric
separation of the electric and heat flows facilitates special
functions. For example, thermoelectric cooling and engine can be
realized using a single type of carrier doping {(i.e., without
  serial connection between $n$- and $p$-type thermoelectric
  components)}
via transverse thermoelectric effect.\cite{TTE} Recently a related, 
but different, thermoelectric effect is found in mesoscopic
thermoelectrics in three-terminal
geometry.\cite{3t0,ora,dot,cavity,prb1,photon,magnon,segal,hopping,jordan,pn,patent,new,qw}
Researches in this direction is pioneered by
the theory of Edwards {\sl et al.}\cite{Edwards} and the later
experiments.\cite{cam-exp}
The underlying physics is illustrated in Fig.~\ref{fig1} ({see
also Ref.~\onlinecite{prb1}}): excess population of 
phonons can induce an electric current during inelastic transport
processes. Heat and electric flows are geometrically separated since
heat is carried by the phonons {flowing from/into the phonon bath}. This picture can be generalized
to inelastic transport processes assisted by other elementary
excitations, such as {photons\cite{photon}}, electron-hole excitations\cite{dot,cavity} and
magnons.\cite{magnon} Besides the quantum dots (QDs) can be replaced
with any conductors given that the carrier energies at the left and
right conductors are considerably different{, which} can be realized by
two low-dimensional structures (e.g., quantum wells\cite{qw} or wires), or a
barrier,\cite{patent} or a band gap.\cite{pn}

Microscopic analysis\cite{3t0,dot,cavity,prb1,photon,magnon,jordan,pn,patent,new,qw}
indicates that the performance of each individual nano-scale
three-terminal thermoelectric (3T-TE) device is promising. Experiments have
demonstrated the effectiveness of 3T-TE
cooling at submicron scale.\cite{cam-exp} In this work we
focus on 3T-TE systems based on the structure illustrated in Fig.~2.
{This structure was initially proposed by Edwards {\sl et
    al.}\cite{Edwards} and later explored experimentally in
  Ref.~\onlinecite{cam-exp} for cooling of electrons at cryogenic
  temperature. Recently Jordan {\sl et al.} extend the idea to
  thermoelectric engine with layered self-assembled QDs where
  considerable electrical current density could be obtained due to
  contributions from many parallel quantum tunneling
  channels\cite{jordan}. This proposal significantly improve the
  potential for thermoelectric energy harvesting of the original idea
  (An extended idea of replacing the QD layers by quantum wells is
  presented in Ref.~\onlinecite{qw}).
  Here we exploit the same structure of Jordan {\sl et al.} to study
  cooperative effects between the longitudinal and transverse
  thermoelectric powers.}

\begin{figure}[htb]
  \includegraphics[height=11.3cm]{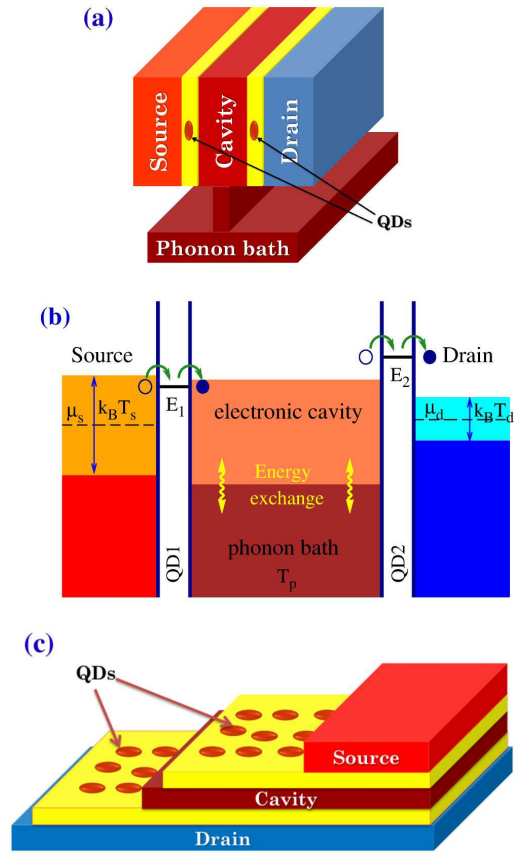}
  \caption{(Color online) Illustration of the 3T-TE device studied
    in this work. (a) Schematic of a possible realistic set-up. The
    layered structure consists of three highly conducting regions: the
    source, cavity, and drain. In between them there are two highly
    resisting layers (yellow regions) with very low thermal and
    electrical conductance. A QD is embedded into each resisting layer
    to allow electrical and thermal conduction through it. The cavity is
    connected with a (insulating) phonon bath. The connection allows
    efficient energy exchange between them. (b) Illustration of the
    microscopic processes. Carrier distributions in the source and
    the drain are determined by the chemical potentials and
    temperatures. Electrons transmit through QDs via resonant
    tunneling. When energies of the two QDs, $E_1$ and $E_2$, are
    different, energy exchange between the cavity and the phonon bath
    is necessary for the establishment of the steady state transport. 
    (c) A scheme for building macroscopic devices where many QDs,
    forming 2D arrays, are embedded in the resisting layers.}
  \label{fig2}
\end{figure}

In Fig.~\ref{fig2}(a) the electronic cavity, as well as the source and the drain,
are highly conducting layers made of heavily-doped semiconductors (e.g., heavily-doped
silicon or GaAs). In between those layers, there are two highly 
resisting layers with high thermal and electrical resistance. Each of
the resisting layer is embedded with a QD through which electrons can
tunnel between the cavity and the electrodes. The resonant tunneling
through QDs are responsible for the transport. The energy levels in
the left and right QDs are $E_1$ and $E_2$, respectively. When $E_1\ne
E_2$, an electron transmitting from the source to the drain takes a
finite amount of energy, $E_2-E_1$, from the cavity. To reach steady
states the cavity must exchange energy with the phonon bath [see
Fig.~\ref{fig2}(b)]. The transverse thermoelectric effect is manifested as
the fact that an electric current drives a heat current from
the phonon bath [see Fig.~\ref{fig2}(b)], and vice versa.

{Following Ref.~\onlinecite{jordan}}, the scheme to assemble the nano-scale devices into a macroscopic
device is straightforward: 2D arrays of QDs can be placed in the
resisting layers [see Fig.~\ref{fig2}(c)]. This can be realized by
self-assembled QDs grown on the surface of
semiconductors,\cite{ass-qd} or as we proposed here, core-shell QDs
embedded in (undoped) polymers with low thermal and electrical
conductance.\cite{polymer,poly,poly2,note1} In such an assembly scheme
the total electric
current is the sum of the electric currents in each 
nano-scale 3T-TE device (i.e., a pair of QDs). High power density can
be prompted by high density of QDs which can reach $\sim
10^{11}$~cm$^{-2}$ for a single layer (about 1~nm
thick).\cite{ass-qd}

To date all works on 3T-TE systems focus on exploiting the transverse
thermoelectric effect. However, there is
also a longitudinal thermoelectric effect in the system: the
temperature difference between the two electrodes can also induce an
electric current. A full description of thermoelectric transport in
3T-TE systems is given by the phenomenological equation (similar
equations were found in Refs.~\onlinecite{ora,prb1,pn,hopping})
\begin{equation}
\left( \begin{array}{c}
    I \\ I_{Q1}\\ I_{Q2} \end{array}\right) =
\left( \begin{array}{cccc}
    G & L_1 & L_2 \\
    L_1 & K_1 & K_{12} \\
    L_2 & K_{12} & K_2 \\
  \end{array}\right) \left(\begin{array}{c}
    V \\ \Delta T_1/T \\ \Delta
    T_2/T \end{array}\right) , \label{3t-trans}
\end{equation}
where $L_1$ and $L_2$ represent the longitudinal and transverse
thermoelectric effects, respectively. $I_{Q1}$ and $I_{Q2}$ stand for
the heat currents leaving the source and the phonon bath,
respectively. The two temperature differences are 
$\Delta T_1=T_{s} - T_d$ and $\Delta T_2 = T_p - T_d$ 
with $T_{s}$, $T_d$, and $T_p$ being the temperatures of the
source, the drain, and the phonon bath, respectively.

In this work we show that, if both the longitudinal and transverse
thermoelectric effects are exploited simultaneously, due to 
cooperation between the two, the efficiency and power can be considerably
improved. {The cooperative effect originates deeply from the
  nature of three-terminal thermoelectric systems: the two
  thermoelectric effects are correlated with each other, or in
  other words, the electrical current are {\em simultaneously} induced by
  the two {\em different} temperature gradients.
Simplified geometric interpretation} is that the electric currents induced
by the two thermoelectric effects can be parallel or anti-parallel. In
the former case the two effects add up constructively, leading to
enhanced thermopower and efficiency. The phenomenon reflects the
cooperative effect of two (or more) correlated thermoelectric effects
which is referred to as the ``cooperative thermoelectric effect''.

The cooperative thermoelectric effect is manifested in the proposed
thermoelectric device. Using  material parameters 
we calculate the thermoelectric transport coefficients as well as the
figure of merit and power factor. It is found that both the figure of
merit and the power factor are considerably improved by the cooperative
thermoelectric effect. {This enhancement is as effective for good
thermoelectrics as that for bad thermoelectrics. In calculation we
show that the enhancement induced cooperative effect changes only
slightly when phonon heat conductivity or QD energy broadening is
increased significantly. These results demonstrate that cooperative
effect is an alternative way to improve the performance of
thermoelectrics in three-terminal geometry effectively.}

This paper is organized as follows: In Sec.~II we establish
thermoelectric transport of the 3T-TE systems from microscopic 
theory. In Sec.~III we {demonstrate the cooperative
thermoelectric effect in a geometric way}. In Sec.~IV we calculate the thermoelectric
transport coefficients as well as the figures of merit and power
factors for the longitudinal, transverse, and cooperative
thermoelectric effects for the 3T-TE systems using 
material parameters. We conclude in Sec.~V. Studies in this work {are 
focused on} linear-response steady state transport. {Interesting
nonlinear effects\cite{nonl} could be discussed in future works.}

\section{Microscopic theory of Thermoelectric transport}
{In this section we develop a microscopic theory of 3T-TE transport,
following the formalism of Jordan {\sl et al.}\cite{jordan}}
The system is a layered structure with thickness
$L_{tot}=L_s+L_d+L_c+2L_{qd}$ where $L_s$, $L_d$, $L_c$, and $L_{qd}$
are the thickness of the source, the drain, the cavity, and the
resisting layers, respectively. For realistic
design, $L_s$, $L_d$, and $L_c$ is about one hundred nanometers (nm),
while $L_{qd}$ is on the order of ten nm. The size of 
QDs is a few nm. The level spacing of the QDs, typically on the order
of 100~meV {for core-shell QDs (see Ref.~\onlinecite{gong})}, is much larger than
$k_BT$. We suggest to fabricate serially connected unit structures [the structure in Fig.~\ref{fig2}(a)] to scale
the device up to a fully three-dimensional macroscopic device which can be implemented via
layer-by-layer growth methods. 

The Hamiltonian of the system is written as
\be
H = H_{s} + H_{d} + H_c + H_{qd} + H_{int},
\ee 
where
$H_s$, $H_{d}$, $H_c$, and $H_{qd}$ are the Hamiltonian of the
source, the drain, the cavity, and the QD, respectively.
$H_{\alpha} = \sum_{\vec k} \varepsilon_{\vec k} c^\dagger_{{\vec
    k},\alpha} c_{{\vec k},\alpha}$,
where $\alpha = s, d, c$ denotes the source, the drain, and the
cavity, respectively. $\varepsilon_{\vec k} = \frac{\hbar^2
  k^2}{2m^\ast}$ with $m^\ast$ being the effective mass of the charge
carrier. 
\be
H_{qd} = \sum_{i} E_{i}^{(\ell)} d_{i}^\dagger d_{i} +
\sum_{j} E_j^{(r)} d_j^\dagger d_j,
\ee 
where the index $i$ and $j$
numerate the QDs in the left $({\ell})$ and right $(r)$ resisting layers,
respectively. $H_{int}$ describes hybridization of the QD states
with the states in the source, drain, and cavity, 
\be
H_{int} =
\sum_{\alpha=s,c} \sum_{{\vec k},i} V^{(\ell)}_{i,\alpha, 
  {\vec k}} c^\dagger_{{\vec k},\alpha} d_i + \sum_{\alpha=d,c}
\sum_{{\vec k},j} V^{(r)}_{j,\alpha, {\vec k}} c^\dagger_{{\vec
    k},\alpha} d_j + {\rm H.c.}
\ee
The electric and thermal currents
through the left resisting layer from the source to the cavity are given by\cite{book} 
\begin{subequations}
\begin{align}
I_{e,\ell} &= \frac{2e}{h} \int dE\ t_{\ell}(E) [f_s(E) - f_c(E)] ,\\
I_{Q,\ell} &= \frac{2}{h} \int dE (E-\mu) t_{\ell}(E) [f_s(E) - f_c(E)], 
\end{align}
\label{current}
\end{subequations}
respectively. The factor of two comes from the spin degeneracy of the carriers.
$e$ is the electronic charge. $f_s$ and $f_c$ are the carrier distribution
functions of the source and the cavity, respectively. They are
determined by the temperatures of the source $T_{s}$ and the cavity
$T_c$ as well as by their electrochemical potentials $\mu_{s}$ and
$\mu_c$. Note that because the voltage and temperature gradients are
mainly distributed at the two resisting layers, as a good
approximation, one can assign uniform chemical potentials and
temperatures to the source, drain, and cavity regions. The
energy dependent transmission through QDs is given
by\cite{book,ora} $t_{\ell}(E) =  \sum_i t_i(E)$ with
\begin{equation}
t_i(E) = \frac{ \hbar^2\Gamma_{s,i}(E) \Gamma_{c,i}^{(\ell)}(E)}{(E-E_i^{(\ell)})^2 + \frac{\hbar^2}{4}
  \left(\Gamma_{s,i}(E)+\Gamma_{c,i}^{(\ell)}(E)\right)^2} ,
\end{equation}
where
\begin{subequations}
\begin{align}
& \Gamma_{s,i}(E) = \frac{2\pi}{\hbar} \sum_{\vec k}
|V^{(\ell)}_{i,s,{\vec k}}|^2\delta( E - \varepsilon_{\vec k}) , \\
& \Gamma_{c,i}^{(\ell)}(E) = \frac{2\pi}{\hbar} \sum_{\vec k} 
|V^{(\ell)}_{i,c,{\vec k}}|^2 \delta(E - \varepsilon_{\vec k}) ,
\end{align} 
\end{subequations}
are the energy-dependent tunneling rates from the QD $i$ to the source 
and the cavity, respectively. The electric and thermal currents from
the drain to the cavity can be  obtained by the
replacements: $i\to j$, $\ell \to r$, and $s\to d$. Introducing
\begin{subequations}
\begin{align}
G_\ell & =  \frac{2e^2}{hk_BT} \int dE\ t_{\ell}(E)  f(E) [1-f(E)] \\
L_\ell & =  \frac{2e}{hk_BT} \int dE (E-\mu) t_{\ell}(E)  f(E)
[1-f(E)]  \\
K_\ell & =  \frac{2}{hk_BT} \int dE (E-\mu)^2 t_{\ell}(E)  f(E)
[1-f(E)] 
\end{align}
\label{GLK}
\end{subequations}
with $f(E)$ being the equilibrium carrier distribution, in the
linear-response regime one can rewrite
Eq.~(\ref{current}) as
\begin{subequations}
\begin{align}
I_{e,\ell} &= G_\ell (\mu_{s} - \mu_c)/e + L_\ell (T_{s} - T_c)/T ,\\
I_{Q,\ell} &= L_\ell (\mu_{s} - \mu_c)/e + K_\ell (T_{s} - T_c)/T .
\end{align}
\label{c22}
\end{subequations}
Expressions for the currents from the drain to the cavity
$I_{e,r}$ and $I_{Q,r}$ can be obtained from the above by the
replacements $\ell \to r$ and $s\to d$.

{Inelastic scatterings, such as the electron-phonon and electron-electron scatterings,
are crucial for the establishment of steady states in the cavity. In
the concerned temperature range, $300\sim 500$~K, those scatterings
are quite efficient. The heat transfer between the phonon bath and the
cavity can be made efficient by using materials with high thermal
conductivity to connect them. Interface thermal resistance can be
reduced if the cavity and the phonon bath are made of the same
material. We assume the thermal conduction between the phonon bath and
the cavity is efficient and the temperature gradient across them and
within the cavity is considerably smaller than that across the two
polymer layers.} In this way the temperature of the cavity is very
close to that of the phonon bath and one can approximate that
$T_c=T_p$.\cite{prb1,pn}

Energy conservation gives $I_{Q,\ell} +
I_{Q,r} + I_{Q,t} + IV = 0$, where $I_{Q,t}$ is the heat current from
the phonon bath to the cavity. Therefore there are only two
independent heat currents.\cite{prb1,hopping} In Eq.~(\ref{3t-trans})
the two independent heat currents are chosen as $I_{Q1}=I_{Q,\ell}$
and $I_{Q2}=I_{Q,t}$. Charge conservation,
$I_{e,\ell}+I_{e,r}=0$, determines the electrochemical potential
of the cavity 
\bea
\mu_c &=& (G_{\ell}+G_r)^{-1} \Big\{ G_\ell \mu_s + G_r
\mu_d + e T^{-1}\big[ L_\ell \left(T_s - T_p\right) \nn\\
&& \ \ +  L_r \left(T_d - T_p\right)\big] \Big\} . 
\eea
Inserting the above into Eqs.~(\ref{3t-trans}) and (\ref{c22}) we obtain
\begin{subequations}
\begin{align}
& G = ( G_\ell^{-1} + G_r^{-1} )^{-1} ,\quad  L_1 = G
\frac{L_\ell}{G_{\ell}} ,\\
& L_2=G(\frac{L_r}{G_r} - \frac{L_\ell}{G_{\ell}}) , 
\quad K_1 = K_\ell - \frac{L_\ell^2}{G_\ell +G_r} ,\\
& K_{12} = \frac{L_\ell (L_\ell + L_r) }{G_\ell + G_r}-K_\ell , \\
& K_2 = K_\ell + K_r - \frac{ (L_\ell + L_r)^2 }{G_\ell + G_r} .
\end{align}
\end{subequations}
To understand these results, we rewrite the transport coefficients
in terms of average electronic energies, following Mahan and Sofo,\cite{ms}
\begin{subequations}
\begin{align}
L_1 &= e^{-1} G \left\langle E-\mu\right\rangle_\ell , \ \ L_2 =
e^{-1} G (\left\langle E\right\rangle_r - \left\langle E\right\rangle_\ell ), \\
K_1 &= \frac{L_1^2}G + K_{tl},\quad K_{12} = \frac{L_1 L_2}G - K_{tl} , \\
K_2 &= \frac{L_2^2}G + K_{tl} + K_{tr} .
\end{align}
\label{ave1}
\end{subequations}
where
\begin{subequations}
\begin{align}
K_{tl} &= e^{-2} G_\ell \Big[\left\langle (E-\mu)^2\right\rangle_\ell
  - \left\langle E-\mu \right\rangle^2_\ell\Big] , \\
K_{tr} &= e^{-2} G_r \Big[ \left\langle (E-\mu)^2\right\rangle_r
  -\left\langle E-\mu\right\rangle^2_r \Big] .
\end{align}
\label{ave2}
\end{subequations}
The average in the above is defined as 
\begin{equation}
\left\langle E^n\right\rangle_\beta \equiv \frac{\int dE E^n G_\beta(E)}{\int dE G_\beta(E)} \label{ave3}
\end{equation}
with $G_\beta(E)=\frac{2e^2}{hk_BT}
t_\beta(E) f(E) [1-f(E)]$, for $\beta=\ell, r$ and
$n=0,1,2$. $G_{\ell}=\int dE G_\ell(E)$ and $G_r=\int dE G_r(E)$.
One readily notices from Eq.~(\ref{ave2}) that $K_{tl}$ and $K_{tr}$ must
be non-negative.

For a macroscopic system with area $A$ the electrical conductivity is
$\sigma=Gl_{u}/A$ with $l_u=2l_{qd}+l_c+(l_s+l_d)/2$ being the
thickness of an unit structure [the structure in Fig.~\ref{fig2}(a)]. 
Similarly the thermal conductivities
are $\kappa_1=K_1l_u/(AT)$, $\kappa_2=K_1l_u/(AT)$, and
$\kappa_{12}=K_{12}l_u/(AT)$. The longitudinal and transverse
thermopowers are
\begin{equation}
  S_1\equiv \frac{L_1}{TG}=\frac{\left\langle
      E-\mu\right\rangle_\ell}{eT}, \ \ S_2\equiv
  \frac{L_2}{TG}=\frac{\left\langle E\right\rangle_r - \left\langle
      E\right\rangle_\ell}{eT} .
\end{equation}
$S_2$ is proportional to the energy difference, reflecting that it is
associated with the inelastic processes. In contrary $S_1$ remains
finite when inelastic processes vanish.

The total entropy production of the system in the
linear response regime is written as 
\begin{equation}
\frac{d S}{dt} = \frac{1}{T}\left( I V + I_{Q1}\frac{\Delta
    T_1}{T} + I_{Q2}\frac{\Delta T_2}{T}\right) .
\end{equation} 
The second law of thermodynamics, $\frac{dS}{dt}\ge 0$, requires that\cite{onsager} 
\begin{align}
& GK_1\ge L_1^2,\ \ GK_2\ge L_2^2,\ \ K_1K_2\ge K_{12}^2, \label{stab}
\end{align}
as well as that the determinant of the $3\times 3$ transport matrix in
Eq.~(\ref{3t-trans}) to be non-negative. Those requirements are
satisfied for the transport coefficients in Eq.~(\ref{ave1}) because
$K_{tl}, K_{tr}\ge 0$.

\section{Cooperative effect: A geometric interpretation}
\label{ana-coop}
We parametrize the two temperature differences as
\begin{equation}
\Delta T_1 \equiv \Delta T \cos\theta, \quad \Delta T_2 \equiv
\Delta T \sin\theta .
\end{equation}
The {exergy efficiency (or the ``second-law
efficiency'', see Refs.~\onlinecite{2nd,general})} of the thermoelectric engine is\cite{2nd,Odum,general}
\be
\phi = \frac{-IV}{I_{Q1}\Delta T_1/T + I_{Q2}\Delta T_2/T} \le
\phi_{max} = \frac{\sqrt{ZT+1}-1}{\sqrt{ZT+1}+1}. \nn
\ee
{The exergy efficiency (``second-law efficiency'') is defined by
  the output {\em free energy} divided by the input {\em free
    energy}.\cite{2nd,Odum,general} It has been widely used in the
  studies of energy conversion in chemical and biological systems since its invention about 60 years
  ago.\cite{Odum} Recently it was applied to thermoelectric
  systems\cite{rev2}. According to Ref.~\onlinecite{onsager} the rate
  of variation of free energy associated with a current is given by
  the product of the current and its conjugated thermodynamic
  force. Hence the denominator of the above equation consists of heat
  currents multiplied by temperature differences. It has been shown in
  Ref.~\onlinecite{general} that the relation between the efficiency
  of $\eta=W/Q$ for heat engine (or $\eta=Q/W$ for refrigerator) and
  the second-law efficiency $\phi$ is that $\phi=\eta/\eta_C$ where
  $\eta_C$ is the Carnot efficiency. Thus Ioffe's figure of merit is
  also obtained starting from the second-law efficiency.}
At given $\theta$ the figure of merit is
\begin{equation}
  ZT = \frac{\sigma S_{\rm eff}^2 T}{\kappa_{\rm eff}-\sigma
    S_{\rm eff}^2 T}  \label{zzt}
\end{equation}
is the figure of merit. Here $S_{\rm eff}=S_1\cos\theta + S_2\sin\theta$ and
$\kappa_{\rm
  eff}=(K_1\cos^2\theta+2K_{12}\sin\theta\cos\theta+K_2\sin^2\theta)l_u/(AT)$. Upon
optimizing the output power of 
the thermoelectric engine, one obtains\cite{maxpower}
\be
W_{max} = \frac{1}{4} P (\Delta T)^2 ,
\ee
with the power factor
\be
P = \sigma S_{\rm eff}^2 . \label{ppower}
\ee
When $\theta=0$ or $\pi$, Eqs.~(\ref{zzt}) and (\ref{ppower}) give the
well-known figure of merit and power factor for the longitudinal
thermoelectric effect\cite{honig,ms} 
\begin{equation}
Z_l T = \frac{\sigma S_1^2 T}{\kappa_1 - \sigma S_1^2 T} , \quad  P_l = \sigma S_1^2 .
\end{equation}
The transverse thermoelectric figure of merit and power factor, i.e.,
$\theta=\pi/2$ or $3\pi/2$, are given by\cite{prb1,pn} 
\begin{equation}
Z_t T = \frac{ \sigma S_2^2 T }{\kappa_2 - \sigma S_2^2 T} , \quad P_t = \sigma S_2^2 .
\end{equation}

\begin{figure}[htb]
  \includegraphics[height=10.4cm]{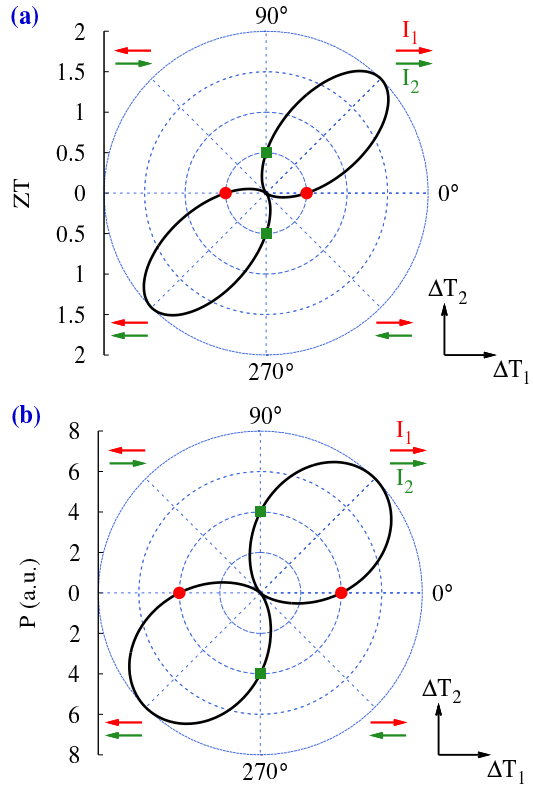}  
  \caption{(Color online) Polar plot of figure of merit $ZT$ (a)
    and power factor $P$ (b) [in arbitrary unit (a.u.)] versus angle
    $\theta$. At $\theta=0^{\circ}$ or $180^{\circ}$ $ZT$ and $P$ recover the
    values for the longitudinal thermoelectric effect (red dots),
    while at $\theta=90^{\circ}$ and $270^{\circ}$ they go back to
    those of the transverse thermoelectric effect (green
    squares). The arrows in the I, II, III, IV quadrants label the
    direction of the currents $I_1\equiv L_1\Delta T_1/T$ (red arrows) and
    $I_2\equiv L_2\Delta T_2/T$ (green arrows). The transport coefficients are:
    $L_1=L_2=2Gk_BT/e$, $K_1=K_2=12G(k_BT/e)^2$, and
    $K_{12}=0$. $I_0\equiv GV$ (not shown) is reversed in III and IV quadrants for the
    operation of the thermoelectric engine.}
\label{fig3}
\end{figure}

Fig.~\ref{fig3}(a) shows $ZT$ versus the angle $\theta$ in a polar plot
for a specific set of transport coefficients satisfying the
thermodynamic bounds in (\ref{stab}). Remarkably for $0<\theta<\pi/2$ and
$\pi<\theta<3\pi/2$, $ZT$ is {\em greater} than both $Z_lT$ and
$Z_tT$. To understand the underlying physics, we decompose the
electric current into three parts $I=I_0+I_1+I_2$ with $I_0\equiv GV$,
$I_1\equiv L_1\Delta T_1/T$, and $I_2=L_2\Delta T_2/T$. The two
thermoelectric effects add up constructively when $I_1$ and $I_2$ have
the same sign which takes place when $0<\theta<\pi/2$ and
$\pi<\theta<3\pi/2$. Fig.~\ref{fig3}(b) shows the power factor versus the angle
$\theta$. The power factor is also {\em larger} when the two currents
$I_1$ and $I_2$ are in the same direction. The cooperation
of the two thermoelectric effects thus leads to enhanced figure of merit
and output power.

One can maximize the figure of merit by tuning the angle
$\theta$. This is achieved at
\be
\partial_\theta (ZT) = 0 .
\ee
We find that the figure of merit is maximized at $\theta=\theta_M$ with
\begin{equation}
\tan(\theta_M) \equiv \frac{ L_2 K_1 - L_1 K_{12} }{ L_1 K_2 - L_2
  K_{12}} = \frac{S_2 \kappa_1 - S_1 \kappa_{12}}{ S_1 \kappa_2 -
  S_2 \kappa_{12} }. \label{thM}
\end{equation}
After some algebraic calculation the maximum figure of merit is found
to be
\begin{equation}
Z_MT = \frac{G(K_1K_2-K_{12}^2)}{ D_{\cal M}} - 1 .
\end{equation} 
where $D_{\cal M}=G K_1 K_2 - G K_{12}^2 -  K_2 L_1^2 + 2 L_1 K_{12} L_2
- K_1 L_2^2 $ denotes the determinant of the $3\times 3$
transport matrix in Eq.~(\ref{3t-trans}). $Z_MT$ is {\em greater} than
{both} $Z_lT$ and $Z_tT$, unless the denominator or the numerator
in Eq.~(\ref{thM}) vanishes. Nevertheless it is guaranteed by
Eqs.~(\ref{ave1}) and (\ref{ave2}) that both the numerator and
denominator in Eq.~(\ref{thM}) is nonzero when the broadening of the
quantum dot energy is finite.

One can also tune $\theta$ to maximize
the power factor $P$ which is achieved at
\be
\partial_\theta P = 0 .
\ee
The power factor is maximized at $\theta=\theta_m$ (in general
$\theta_m\ne \theta_M$) with
\be
\tan(\theta_m) \equiv \frac{L_2}{L_1}=\frac{S_2}{S_1}.
\ee
The maximum power factor 
\be
P_m=\sigma (S_1^2+S_2^2)
\ee 
is greater than {both} $P_l$ and $P_t$ unless $S_1$ or $S_2$ is zero. If
$\theta_M$ is close to $\theta_m$, both the figure of merit and the
power factor can be improved by the cooperative effect simultaneously
in certain range of $\theta$.

The cooperative thermoelectric effect becomes particularly simple and
vivid when $K_{12}=0$, $K_1=K_2\equiv K$, $L_1=L_2\equiv L$, and
$S\equiv L/(TG)$. In this special case the transport equation becomes
\begin{equation}
\left( \begin{array}{c}
    I \\ I_{Q1}\\ I_{Q2} \end{array}\right) =
\left( \begin{array}{cccc}
    G & L & L \\
    L & K & 0 \\
    L & 0 & K \\
  \end{array}\right) \left(\begin{array}{c}
    V \\ \Delta T_1/T \\ \Delta
    T_2/T \end{array}\right) .
\end{equation}
We shall use the following combinations of temperature differences
\be
\Delta T_a = \frac{\Delta T_1 + \Delta T_2}{\sqrt{2}}, \quad \Delta
T_b = \frac{\Delta T_1 - \Delta T_2}{\sqrt{2}} .
\ee
The heat currents conjugate to the above forces are
\be
I_{Qa} = \frac{ I_{Q1} + I_{Q2} }{\sqrt{2}}, \quad  I_{Qb} = \frac{
  I_{Q1} - I_{Q2} }{ \sqrt{2} } .
\ee
The transport equation then becomes
\begin{equation}
\left( \begin{array}{c}
    I \\ I_{Qa}\\ I_{Qb} \end{array}\right) =
\left( \begin{array}{cccc}
    G & \sqrt{2}L & 0 \\
    \sqrt{2}L & K & 0 \\
    0 & 0 & K \\
  \end{array}\right) \left(\begin{array}{c}
    V \\ \Delta T_a/T \\ \Delta
    T_b/T \end{array}\right) .
\end{equation}
Consider conversion of heat $I_{Qa}$ into work $-IV$ at $\Delta T_a\ne
0$ and $\Delta T_b=0$. %The efficiency is $\eta=-IV/I_{Qa}$. 
The figure of merit and power factor are
\begin{equation}
ZT=\frac{2L^2}{GK-2L^2}=\frac{2\sigma S^2T}{\kappa - 2\sigma S^2T},
\quad P=2 \sigma S^2 , 
\end{equation}
respectively, with $\kappa\equiv Kl_u/(AT)$. The above figure of merit and power factor are {\em
  greater} than those of the longitudinal and transverse
thermoelectric effects which are
\be
Z_lT = Z_tT = \frac{\sigma S^2T}{\kappa - \sigma S^2T}, \quad P_l = P_t
= \sigma S^2 .
\ee

\section{Calculation of thermoelectric performance}
\label{sec:robust}

\begin{figure}[htb]
  \includegraphics[height=12.0cm]{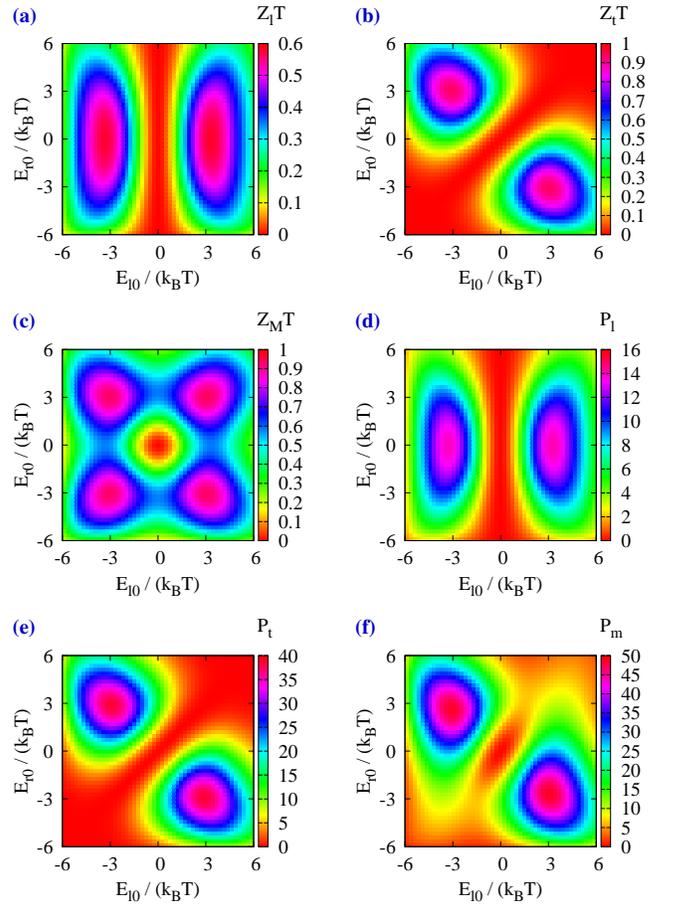}
  \caption{{(Color online) Figures of merit $Z_lT$ (a), $Z_tT$ (b), and
    $Z_MT$ (c), and power factors (in unit of
    $10^{-4}$~W~m$^{-1}$~K$^{-2}$) $P_l$ (d), $P_t$ (e), and $P_m$ (f)
    versus the QD energy $E_{\ell 0}$ and $E_{r0}$ for
    $\Delta_{qd}=20$~meV and $\kappa_p = 0.05$~W~m$^{-1}$~K$^{-1}$. The temperature is
    $T=400$~K, and the sheet density of QDs in each polymer
    layer (thickness 20~nm) is $4\times 10^{12}$~cm$^{-2}$. QD
    tunneling linewidth is $\Gamma=$30~meV.} }
\label{fig4}
\end{figure}

We now calculate the transport coefficients using  material 
parameters. The robustness of the device
performance is tested by including the randomness of QD
energy. Beside the transport mechanism described in
Sec.~II, there are other mechanisms conducting heat among
reservoirs. The most important one is the heat conduction across the
resisting layers by phonons. Polymers are good thermal
insulators with heat conductivity 
$0.05\sim 1$~W~m$^{-1}$~K$^{-1}$.\cite{polymer,poly,poly2} 
Phonon thermal conductivity should be much reduced in the nano-scale thin
films concerned here due to abundant scattering with the embedded QDs and
interfaces.\cite{nano1} We take the phonon thermal conductivity as
$\kappa_p = 0.05$~W~m$^{-1}$~K$^{-1}$ (e.g., bulk rayon\cite{polymer} has
thermal conductivity of 0.05~W~m$^{-1}$~K$^{-1}$). The thermal
conductance across a resisting layer is $K_p= A \kappa_p T /
l_{qd}$. Adding this contribution to the transport equation and using
Eq.~(\ref{ave1}) leads to 
\begin{align}
& K_1 = \frac{L_1^2}G + K_{tl} + K_p,\quad K_{12} = \frac{L_1L_2}G - K_{tl} - K_p ,\nn\\
& K_2 = \frac{L_2^2}G + K_{tr} + K_{tl} + 2K_{p} .
\end{align}
The variance of the QDs energy is several
tens of meV as revealed by experiments.\cite{ass-qd} The thickness of
the resisting layer is taken as $l_{qd}=20$~nm. The thickness of the
source, the drain, and the cavity are all equal to 80~nm. High QD
density, $2\times 10^{13}$~cm$^{-2}$, has been realized in polymer
matrices of thickness around 20~nm.\cite{poly} It is favorable to have
more than one layer of QDs in each resisting layer to enable
sufficient electron tunneling. Serial tunneling through several QDs
may happen in the transmission across the resisting
layers {with many QDs}.\cite{Gurvitz} We suggest to incorporate $\simeq 4\times
10^{12}$~cm$^{-2}$ QDs into a 20~nm polymer layer which corresponds to
a volume density of QDs as $\simeq 2\times 10^{18}$~cm$^{-3}$ (average
inter-dot distance $\simeq 8$~nm). Taking into account of finite QD
size, the inter-dot tunneling linewidth is
{$1\sim 30$~meV depending on materials and structures
which is taken as $30$~meV here (see experimental measured value of
30~meV in Ref.~\onlinecite{awsh}). And we take $\Gamma\equiv
\Gamma_{c,i}^{(\ell)}(E)=\Gamma_{c,j}^{(r)}(E)=\Gamma_{s,i}(E)=\Gamma_{d,j}(E)=30$~meV 
($\forall i, j, E$) as the QD tunneling linewidth used in calculating
the tunneling rate.} We shall study how the performance of the device
varies with the variance and the mean value of the energy of QDs. The
random QD energy in the left (right) resisting layer is modeled by a Gaussian
distribution centered at $E_{\ell 0}$ ($E_{r 0}$) with a variance $\Delta_{qd}^2$,
\be
g_\beta(E)=\frac{1}{\Delta_{qd}\sqrt{2\pi}}\exp\left[-\frac{(E-E_{\beta
      0})^2}{2\Delta_{qd}^2}\right] .
\ee
with $\beta=\ell, r$. The QD energy and size can be controlled by various
chemical\cite{poly} and physical\cite{ass-qd} methods during growth.
We consider situations with temperature
$T=400$~K. The energy zero is set to be the equilibrium chemical
potential. The band edge of the semiconductor
that constitutes the source, drain, and cavity layers is 200~meV below 
the chemical potential (a typical value for heavily-doped
semiconductors). The electrical conductivity, thermopowers, and thermal
conductivities are calculated according to Eqs.~(\ref{ave1}),
(\ref{ave2}), and (\ref{ave3}). Based on those transport coefficients
we calculate the figures of merit and power factors, $Z_lT$, $Z_tT$,
$Z_MT$ [Figs.~\ref{fig4}(a), \ref{fig4}(b), and \ref{fig4}(c)], $P_l$,
$P_t$, and $P_m$ [Figs.~\ref{fig4}(d), \ref{fig4}(e), and
\ref{fig4}(f)].

{Fig.~\ref{fig4} indicates that the figure of merit $Z_lT$ is optimized at
  the two points $E_{\ell 0}\simeq \pm 3k_BT$ with $E_{r 0}=0$, while
  $Z_tT$ is optimized at $E_{\ell 0}= - E_{r 0}\simeq \pm
  3k_BT$. These results, which are consistent with the results in 
  Ref.~\onlinecite{jordan}, can be understood as the balance between large
  electrical conductivity, large thermopower, and small thermal
  conductivity in optimizing the figure of merit.
  The power factors, $P_l$ and $P_t$, are optimized at parameters
  similar to those of $Z_lT$ and $Z_tT$, respectively.
  We find that when $Z_tT$ is optimized, $Z_MT$ is only slightly
  larger than $Z_tT$. For other situations, $Z_MT$ is considerably
  greater than both $Z_lT$ and $Z_tT$. Particularly near the 
  two points $E_{\ell 0}=E_{r 0}\simeq \pm 3k_BT$, the enhancement of
  figure of merit induced by cooperative effect is
  significant. For the power factor, cooperative effect always leads
  to considerable enhancement of power factor, unless when $P_l$ or
  $P_t$ are close to zero.}

{The above results reveal that cooperative effects can effectively
  improve the figure of merit and power factor for thermoelectrics in
  three-terminal geometry. Such improvement is especially useful
  for systems of which the electronic structure has not been fully
  optimized. Hence cooperative effects offer an additional way to
  improve the performance of thermoelectrics that are potentially useful for
  realistic systems. We also note that the largest figure of
  merit and power factor for the longitudinal thermoelectric effect
  are $Z_lT=0.6$ and $P_l=1.6\times 10^{-3}$~W~m$^{-1}$~K$^{-2}$
  respectively, while the largest figure of merit and power factor for
  the transverse thermoelectric effect is $Z_tT=1$ and $P_t=4\times
  10^{-3}$~W~m$^{-1}$~K$^{-2}$ respectively. This result confirms the
  conclusion in Refs.~\onlinecite{prb1,pn,hopping,jordan,qw} that 
  the transverse thermoelectric effect in three-terminal geometry is
  of potential advantages.}

\begin{figure}[htb]
  \includegraphics[height=12.0cm]{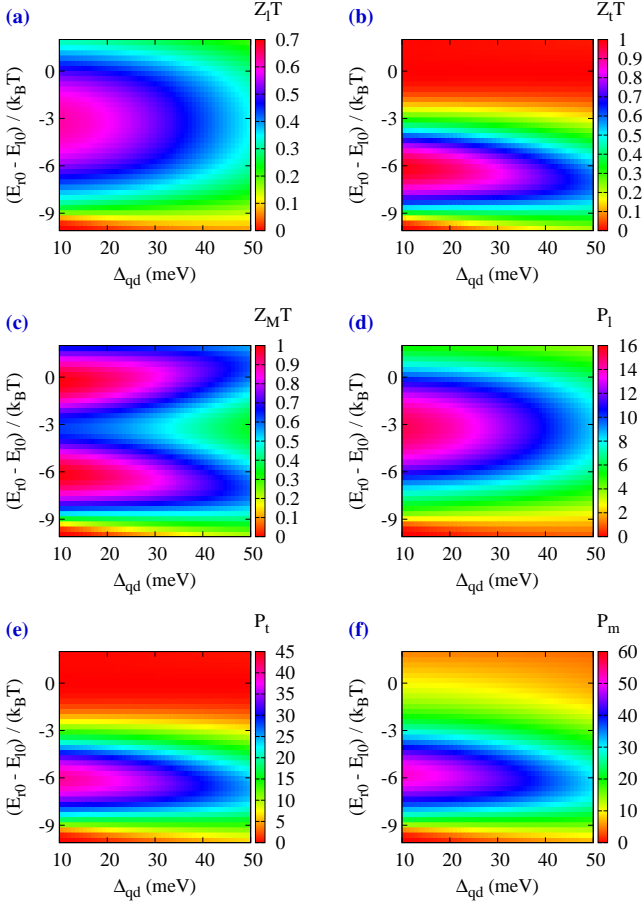}
  \caption{(Color online) Figures of merit $Z_lT$ (a), $Z_tT$ (b), and
    $Z_MT$ (c), and power factors (in unit of
    $10^{-4}$~W~m$^{-1}$~K$^{-2}$) $P_l$ (d), $P_t$ (e), and $P_m$ (f)
    versus the variance of the QDs energy $\Delta_{qd}$ and the difference
    of the mean values of QD energy in the two resisting layers
    $E_{r0}-E_{\ell 0}$. Here $E_{r0}$ is varying while {$E_{\ell
        0}=110$~meV ($\simeq 3k_BT$)} is fixed. The temperature is
    $T=400$~K, {$\kappa_p = 0.05$~W~m$^{-1}$~K$^{-1}$}, and the
    sheet density of QDs in each polymer 
    layer (of thickness 20~nm) is $4\times
    10^{12}$~cm$^{-2}$. QD tunneling linewidth is $\Gamma=30$~meV. }
\label{fig5}
\end{figure}

{In order to check the robustness of the effect in realistic
  situations we discuss the effect of the broadening of QD energy
  $\Delta_{qd}$ and the energy difference $E_{r0}-E_{\ell 0}$ for 
$E_{\ell 0}=110$~meV ($\simeq 3k_BT$). {Increase of the broadening of QD
  energy $\Delta_{qd}$ reduces the thermopower and increases the
  electronic heat conductivity. Therefore the figures of merit and
  power factors for the longitudinal, transverse, and cooperative
  thermoelectric effects are all reduced.} From Fig.~\ref{fig5} one finds
that considerably large figures of merit and power factors can still
be obtained for broadening of QD energy up to 50~meV.} In experiments
the full width at half-maximum of photoluminescence spectra can be as
small as 35~meV (i.e., the variance is 15~meV).\cite{ass-qd} Thus the
proposed device is of potential application values. Finally in the
above discussions only one energy level in each QD is considered. 
Careful calculation with higher energy levels (100~meV higher)
included indicates that the figure of merit and the power factor are
even larger {[see Appendix]}.

\begin{figure}[htb]
  \centerline{\includegraphics[height=12.0cm]{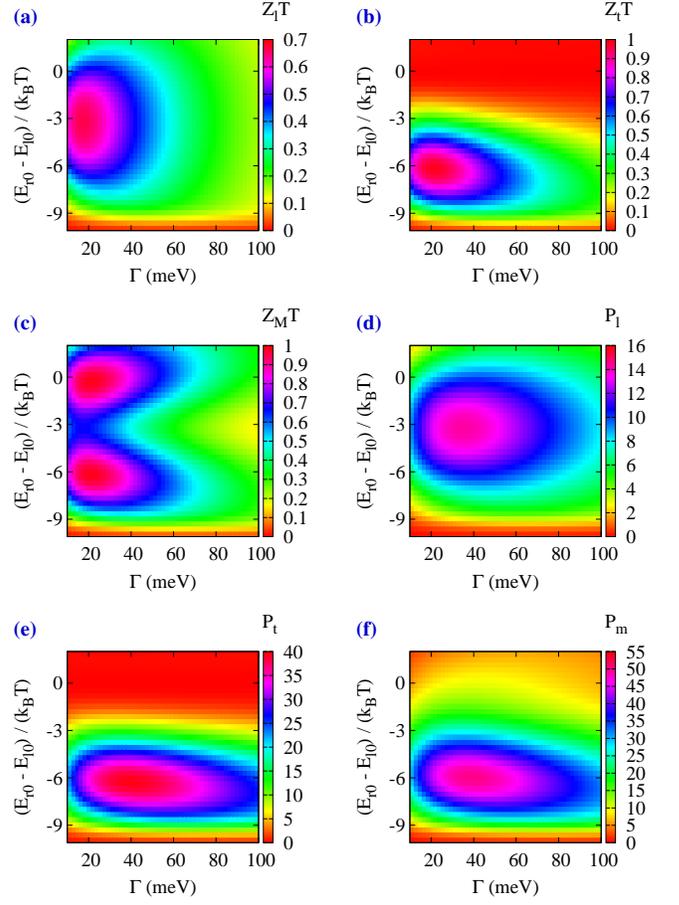}}
  \caption{ (Color online) {Figures of merit $Z_lT$ (a), $Z_tT$ (b), and
    $Z_MT$ (c), and power factors (in unit of
    $10^{-4}$~W~m$^{-1}$~K$^{-2}$) $P_l$ (d), $P_t$ (e), and $P_m$ (f)
    versus the tunneling linewidth of QDs $\Gamma$ and the difference
    of the mean values of QD energy in the two resisting layers
    $E_{r0}-E_{\ell 0}$. Here $E_{r0}$ is varying while {$E_{\ell
        0}=110$~meV ($\simeq 3k_BT$)} is fixed. The temperature is
    $T=400$~K, {$\kappa_p = 0.05$~W~m$^{-1}$~K$^{-1}$}, and the
    sheet density of QDs in each polymer 
    layer (of thickness 20~nm) is $4\times
    10^{12}$~cm$^{-2}$. The variance of the QDs energy is $\Delta_{qd}=20$~meV.} }
 \label{figs}
\end{figure}

{We also plot the figures of merit and power factors for
  longitudinal, transverse, and cooperative thermoelectric effects
  as functions of the tunneling linewidth of QDs $\Gamma$ and the
  energy difference $E_{r0}-E_{\ell 0}$ for $E_{\ell 0}=110$~meV
  ($\simeq 3k_BT$) in Fig.~\ref{figs}. Unlike the monotonic dependence
  on the variance of the QDs energy $\Delta_{qd}$, the figures of
  merit and power factors first increases and then decreases with
  increasing $\Gamma$. This behavior is because at small $\Gamma$
  increase of $\Gamma$
  enhances electron tunneling and hence improves the electrical
  conductivity and the power factors. The enhancement of electron
  tunneling also improves electronic heat conductivity and reduces the
  effect of phonon heat conductivity on the figures of
  merit. Therefore, the figures of merit of the longitudinal,
  transverse, and cooperative thermoelectric effects are improved as
  well. However, the tunneling linewidth $\Gamma$ also induces
  broadening of the energy of transported electron. When such
  broadening is comparable with or larger than the thermal energy
  $k_BT$, it considerably reduces the thermopowers and increases the
  electronic thermal conductivity, hence the power factors $P_l$,
  $P_t$, and $P_m$ as well as the figures of merit $Z_lT$, $Z_tT$, and
  $Z_MT$ are reduced. The optimal tunneling
  linewidths for $Z_lT$, $Z_tT$, and $Z_MT$ are $\Gamma=17.3$, 21.4,
  21.4~meV respectively where the optimal figures of merit are
  $Z_lT=0.668$, $Z_tT=0.987$, and $Z_MT=0.989$ respectively. Besides,
  the optimal tunneling linewidths for $P_l$, $P_t$, and
  $P_m$ are $\Gamma=34.5$, 39.4, and 38.6~meV respectively
  where the optimal power factors are $P_l=1.43\times
  10^{-3}$~W~m$^{-1}$~K$^{-2}$, $P_t=3.96\times
  10^{-3}$~W~m$^{-1}$~K$^{-2}$, and $P_m=5.02\times
  10^{-3}$~W~m$^{-1}$~K$^{-2}$  respectively. The optimal figure of
  merit and power factor for the cooperative thermoelectric effect are
  larger than those of the transverse and longitudinal thermoelectric
  effects. This result is consistent with the proof in
  Sec.~\ref{ana-coop} that the figure of merit and the power factor of
  the cooperative thermoelectric effect is greater than or equal to
  those of the longitudinal and transverse thermoelectric effects.
  Meanwhile the optimal performance of the transverse
  thermoelectric effect is also better than that of the longitudinal
  thermoelectric effect. Overall there are more parameter
  regions for the cooperative thermoelectric effect to have large
  values of the figure of merit and the power factor.}

\begin{figure}[htb]
  \centerline{\includegraphics[height=8.0cm]{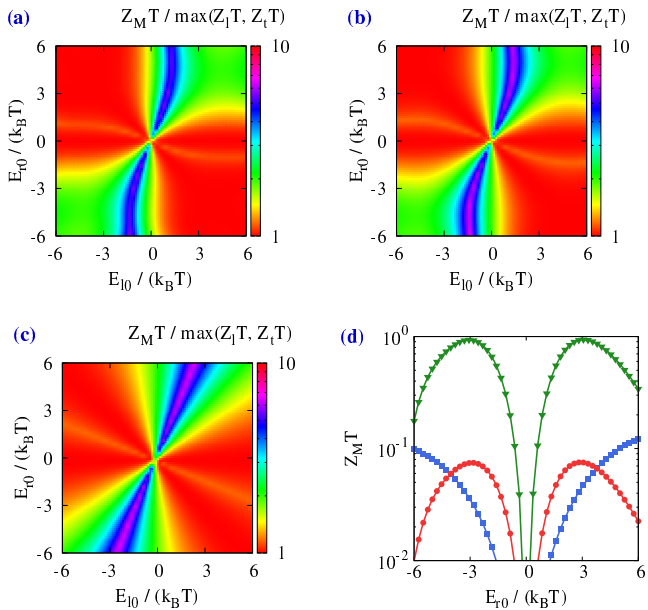}}
  \caption{ (Color online) {The enhancement of figure of merit
      $Z_MT/{\rm max}(Z_lT,Z_tT)$ versus the QD energy $E_{\ell 0}$ and
      $E_{r0}$ for $\kappa_p = 0.05$~W~m$^{-1}$~K$^{-1}$ and
      $\Delta_{qd}=20$~meV (a), as well as when the phonon heat
      conductivity is increased to $\kappa_p = 1$~W~m$^{-1}$~K$^{-1}$
      or when the QD energy broadening is increased to
      $\Delta_{qd}=150$~meV (c). In (d) figures of merit $Z_MT$ as a
      function of $E_{r0}$ with $E_{\ell 0}=E_{r0}$ is plotted for
      $\kappa_p = 0.05$~W~m$^{-1}$~K$^{-1}$ and $\Delta_{qd}=20$~meV
      (curve with triangles), and for when the phonon heat
      conductivity is increased to $\kappa_p = 1$~W~m$^{-1}$~K$^{-1}$
      (curve with dots) or when the QD energy broadening is
      increased to $\Delta_{qd}=150$~meV (curve with squares). The
      temperature is $T=400$~K, and the sheet density of QDs in each
      polymer layer (thickness 20~nm) is $4\times
      10^{12}$~cm$^{-2}$. QD tunneling linewidth is $\Gamma=30$~meV.} }
 \label{fig6}
\end{figure}

{Finally we demonstrate the robustness of the cooperative effects by
examining the enhancement factor $Z_MT/{\rm max}(Z_lT,Z_tT)$ of
thermoelectric figure of merit for $\kappa_p =
0.05$~W~m$^{-1}$~K$^{-1}$ and $\Delta_{qd}=20$~meV, as well as when
the phonon parasitic heat conductivity is increased to
1~W~m$^{-1}$~K$^{-1}$ or when the QD energy broadening is increased to
$\Delta_{qd}=150$~meV. The results are plotted for different QD
energies in Fig.~\ref{fig6}. Considerable enhancement of figure of
merit by cooperative effect is found around $E_{\ell 0}\sim E_{r0}$
for a large portion of parameter region. Moreover, this enhancement by
cooperative effect is still effective when the phonon heat
conductivity is much enhanced or when the QD energy broadening is
significantly increased. This result reveals that the cooperative
effect remains effective in improving thermoelectric efficiency even
in systems with small figure of merit induced by significant parasitic
heat conductivity [see Fig.~\ref{fig6}(d)].}

\section{Conclusion and discussions}

In summary we propose to enhance the thermoelectric efficiency and
power by exploiting cooperative effects in three-terminal
geometry. The three terminal geometry enables {\em one} electric current to
couple with {\em two} temperature gradients with the help of inelastic transport
processes. A scheme exploiting quantum-dots embedded in polymer matrices in
multiple-layered structures is suggested to realize the
principle.  According to calculations based on  material
parameters, the figure of merit and power factor of the proposed
structure are high, which indicates that layered resonant tunneling
structures are potentially good thermoelectric
systems.\cite{layered,rtd} Marked improvements of figure of merit and
power factor by the cooperative thermoelectric effect are
obtained. Remarkably {the enhancement of figure of merit and power
  factor induced by cooperative effects is robust to the parasitic
  phonon heat conductivity as well as quantum dots energy broadening.
  Hence we shown that cooperative effect offers
  an effective way to improve the figure of merit and power factor for
  three-terminal thermoelectric systems, particularly useful for
  systems of which the electronic structure has not been optimized.} Study in
this work indicates that exploiting geometric aspect, inelastic
processes, and cooperative thermoelectric effects could provide
alternative routes to high performance thermoelectrics.

\section*{Acknowledgements}
We thank Baowen Li, Dvira Segal, Ming-Qi Weng, and Daoyong Chen for
illuminating discussions. This work was partly supported by the NSERC
and CIFAR of Canada, and the National Natural Science Foundation of
China Grant No. 11334007.

\begin{appendix}

\section*{Appendix: Effects of higher levels in QDs on thermoelectric properties}

\begin{figure}[htb]
  \includegraphics[height=12.0cm]{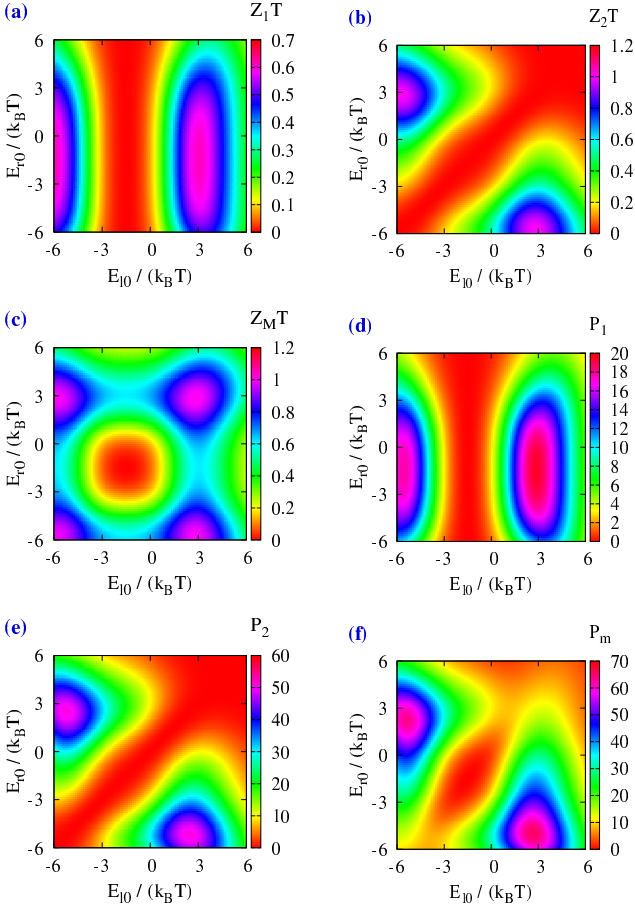}
  \caption{{(Color online) Figures of merit $Z_lT$ (a), $Z_tT$ (b), and
      $Z_MT$ (c), and power factors (in unit of
      $10^{-4}$~W~m$^{-1}$~K$^{-2}$) $P_l$ (d), $P_t$ (e), and $P_m$ (f)
      versus the QD energy $E_{\ell 0}$ and $E_{r0}$ when there is
      another energy level of 100~meV higher than the original one and
      of the same tunneling rate in each QD for both the left and the right
      polymer layers. $\Delta_{qd}=20$~meV and $\kappa_p =
      0.05$~W~m$^{-1}$~K$^{-1}$. The temperature is $T=400$~K, and the
      sheet density of QDs in each polymer 
      layer (thickness 20~nm) is $4\times 10^{12}$~cm$^{-2}$. QD tunneling linewidth is $\Gamma=30$~meV.} }
  \label{figs1}
\end{figure}

\begin{figure}[htb]
  \includegraphics[height=4.0cm]{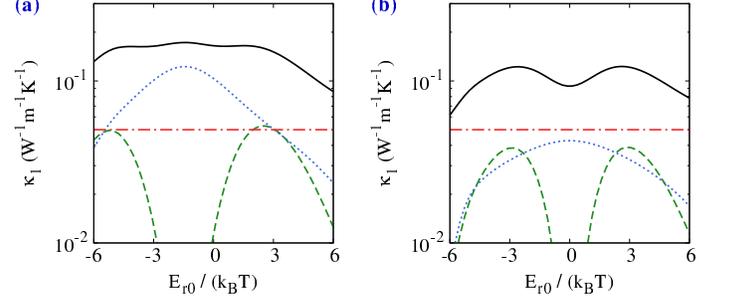}
  \caption{{(Color online) Heat conductivity $\kappa_1$ (black
      full curve) and its three contributions: $\sigma S_1^2 T$ (green
      dashed curve), $[\sigma/(e^2T)]\Big[\left\langle
        (E-\mu)^2\right\rangle_\ell - \left\langle E-\mu
      \right\rangle^2_\ell\Big]$ (blue dotted curve), and $\kappa_p$ (red
      chained curve) as functions of $E_{r0}=E_{\ell 0}$ for (a) with
      the higher level and (b) without the higher level. The higher level is 100~meV higher than
      the original level in each QD. $\Delta_{qd}=20$~meV and $\kappa_p =
      0.05$~W~m$^{-1}$~K$^{-1}$. The temperature is $T=400$~K, and the
      sheet density of QDs in each polymer 
      layer (thickness 20~nm) is $4\times 10^{12}$~cm$^{-2}$. QD tunneling linewidth is $\Gamma=30$~meV.} }
  \label{figs2}
\end{figure}

{We calculate the figures of merit $Z_lT$, $Z_tT$, and
$Z_MT$ and power factors $P_l$, $P_t$, and $P_m$ for the
situation when there is another energy level of 100~meV higher than
the original one in each QD for both the left and the right polymer
layers. {The results are plotted in Fig.~\ref{figs1}.}
We assume that the tunneling rate of the higher level is the
same as the lower one, i.e., $\Gamma=30$~meV. It is seen that the
figures of merit as well as the power factors are all larger when the
higher level is taken into account. Qualitatively, the results here is
similar to those in Fig.~\ref{fig4} but with the center shifted toward
lower energy for both $E_{\ell 0}$ and $E_{r0}$. This observation
reveals that the main effect of the higher level is to enhance the
electrical conductivity as well as the thermopower, which can be
understood easily since a {\em higher} energy channel is introduced. However,
introducing such a channel also increases the thermal conductivity
which normally would reduce the figure of merit. }

{To clarify the underlying mechanism, we plot the thermal conductivity
$\kappa_1$ as a function of $E_{r0}$ when $E_{\ell 0}=E_{r0}$. In
Fig.~\ref{figs2} we plot three different contributions of $\kappa_1$:
$\sigma S_1^2 T$, $[\sigma/(e^2T)]\Big[\left\langle
  (E-\mu)^2\right\rangle_\ell - \left\langle E-\mu
\right\rangle^2_\ell\Big]$, and $\kappa_p$. The figure of merit $Z_lT$
is given by $Z_lT = \sigma S_1^2 T/(\kappa_1-\sigma S_1^2 T)$.
Indeed the thermal conductivity due to energy
uncertainty $[\sigma/(e^2T)]\Big[\left\langle
  (E-\mu)^2\right\rangle_\ell - \left\langle E-\mu
\right\rangle^2_\ell\Big]$ increases when higher level is
introduced. However, in Fig.~\ref{figs2}(a) (with the higher level) when
$\sigma S_1^2 T$ reaches its maximum value, it is very close to
the other two contributions. In comparison, in Fig.~\ref{figs2}(b)
(without the higher level) when $\sigma S_1^2 T$ reaches its maximum
value, it is considerably smaller than the phonon heat conductivity
$\kappa_p$. Therefore, the figure of merit $Z_lT$ is enhanced when the
higher level is taken into account. This particular feature is because
in the case without the higher level, phonon heat conductivity
predominately limits the figure of merit, rather than the variance of
electronic energy.}

\end{appendix}


\begin{thebibliography}{99}
\bibitem{note1} The polymer electronic band gap should be considerably larger than the
  semiconductor band gap, so that the charge carriers in the doped
  semiconductor regions (i.e., the cavity, source, and drain) will not
  diffuse into the polymer resisting   layers.
\end{thebibliography}
\end{document}